\documentclass[aps,prl,twocolumn,showpacs,superscriptaddress,groupedaddress]{revtex4}  
\usepackage{graphicx}  
\usepackage{dcolumn}   
\usepackage{bm}        
\usepackage{amssymb,amsmath,amsopn,amsfonts}
\usepackage{colordvi}
\usepackage{color}

\begin{document}

\widetext


\title{Quantum State Tomography via Linear Regression Estimation}
\author{B. Qi$^{1}$, Z. B. Hou$^{2}$, L. Li$^{2}$, D. Dong$^{3, 4}$, G. Y. Xiang$^{2,\star}$, G. C. Guo$^{2}$\\
$^{1}$\textit{Key Laboratory of Systems and Control, ISS, and National Center for Mathematics and Interdisciplinary Sciences, Academy of Mathematics and Systems Science, CAS, Beijing 100190, P. R. China}\\
$^{2}$\textit{Key Laboratory of Quantum Information,University of Science and Technology of China, CAS, Hefei 230026, P. R. China}\\
$^{3}$\textit{School of Engineering and Information Technology, University of New South Wales at the Australian Defence Force Academy, Canberra, ACT 2600, Australia}\\
$^{4}$\textit{CSC, State Key Laboratory of Industrial Control Technology, Zhejiang University, Hangzhou 310027, P. R. China}\\
$^\star$\textit{email: gyxiang@ustc.edu.cn}}
\date{\today}

\begin{abstract}
A simple yet efficient method of linear regression estimation (LRE) is presented for quantum state tomography. In this method, quantum state reconstruction is converted into
a parameter estimation problem of a linear regression model and the least-squares method is employed to estimate the unknown parameters. An asymptotic mean squared error (MSE) upper bound for all possible states to be estimated is given analytically, which depends explicitly upon the involved measurement bases. This analytical MSE upper bound can guide one to choose optimal measurement sets. The computational complexity of LRE is $O(d^4)$ where $d$ is the dimension of the quantum state. Numerical examples show that LRE is much faster than maximum-likelihood estimation for quantum state tomography.
\end{abstract}

\maketitle


One of the essential tasks in quantum technology is to
verify the integrity of a quantum state \cite{nilsen}. Quantum state tomography has become a standard technology for inferring the state of a quantum system through appropriate measurements and estimation \cite{paris,James16,rehacek,liu,lundeen,salvail,nunn}. To reconstruct a quantum state, one may first perform measurements on a collection of identically prepared copies of a quantum system (data collection) and
then infer the quantum state from these measurement outcomes using appropriate estimation algorithms (data analysis). Measurement on a quantum system generally gives a probabilistic result and an individual measurement outcome only provides limited information on the state of the system, even when an ideal measurement device is used. In principle, an infinite number of measurements are required to determine a quantum state precisely. However, practical quantum state tomography consists of only finite measurements and appropriate estimation algorithms. Hence, the choice of optimal measurement sets and the design of efficient estimation algorithms are two critical issues in quantum state tomography.

Many results have been presented for choosing optimal measurement sets to increase the estimation accuracy and efficiency in quantum state tomography \cite{adamson,Wootters1989,burgh}. Several sound choices that can provide excellent
performance for tomography are, for instance, tetrahedron measurement bases, cube measurement sets, and mutually unbiased bases \cite{burgh}. However, for most existing results, the optimality of a given measurement set is only verified through numerical results \cite{burgh}. There are few methods that can analytically give an estimation error bound \cite{christandl,cramer,zhuhuangjundoctor}, which is essential to evaluate the optimality of a measurement set \cite{DArianoPRL2007,BisioPRL,RoyScott} and the appropriateness of an estimation method.

For estimation algorithms, several useful methods including maximum-likelihood estimation (MLE) \cite{paris,teoPRL,teo,Blume-KohoutPRL,smolin}, Bayesian mean estimation (BME) \cite{paris,huszar,kohout} and least-squares (LS) inversion \cite{opatrny} have been proposed for quantum state reconstruction. The MLE method simply chooses the state estimate that gives the observed results with the highest probability. This method is asymptotically optimal in the sense that the estimation error can asymptotically achieve the Cram\'{e}r-Rao bound. However, MLE usually involves solving a large number of nonlinear equations where their solutions are  notoriously difficult to obtain and often not unique. Recently, an efficient method has been proposed for computing the maximum-likelihood quantum state from measurements
with additive Gaussian noise, but this method is not general \cite{smolin}. Compared to MLE, BME can always give a unique state estimate, since it constructs a state from an integral averaging over all possible quantum states with proper weights. The high computational complexity of this method significantly limits its application. The LS inversion method can be applied when measurable quantities exist that are linearly related to all density matrix elements of the quantum state being reconstructed \cite{opatrny}. However, the estimation result may be a nonphysical state and the mean squared error (MSE) bound of the estimate cannot be determined analytically.

In this Letter, we present a new linear regression estimation (LRE) method for quantum state tomography that can identify optimal measurement sets and reconstruct a quantum state efficiently. We first convert the quantum state reconstruction into a parameter estimation problem of a linear regression model \cite{rao}. Next, we employ an LS algorithm to estimate the unknown parameters. The positivity of the reconstructed state can be guaranteed by an additional least-squares minimization problem. The total computational complexity is $O(d^4)$ where $d$ is the dimension of the quantum state. In order to evaluate the performance of a chosen measurement set,
an MSE upper bound for all possible states to be estimated is given analytically. This MSE upper bound depends explicitly upon the involved measurement bases, and can guide us to choose the optimal measurement set.  The efficiency of the method is demonstrated  by examples on qubit systems.

\emph{Linear regression model.} We first convert the quantum state tomography  problem into a parameter estimation problem of a linear regression model.
Suppose the dimension of the Hilbert space $\mathcal{H}$ of the system of interest is $d$, and $\{\Omega_{i}\}^{d^2-1}_{i=0}$ is a complete basis set of orthonormal operators on the corresponding Liouville space, namely, $\textmd{Tr}(\Omega^{\dag}_i\Omega_j)=\delta_{ij}$, where $\dag$ denotes the Hermitian adjoint and $\delta_{ij}$ is the Kronecker function.  Without loss of generality, let $\Omega_{i}=\Omega_{i}^{\dag}$ and $\Omega_{0}=(1/d)^{\frac{1}{2}}I$, such that the other bases are traceless. That is $\textmd{Tr}(\Omega_{i})=0$, for $i = 1,\ 2,\ \cdots,\ d^2-1$. The quantum state $\rho$ to be reconstructed  may be parameterized as
\begin{equation}\label{rho}
\rho=\frac{I}{d}+\sum^{d^2-1}_{i=1}\Theta_i\Omega_i,
\end{equation}
where $\Theta_i=\textmd{Tr}(\rho\Omega_i)$. Given a set of measurement bases $\{|\Psi\rangle\langle\Psi|^{(n)}\}^{M}_{n=1}$, each $|\Psi\rangle\langle\Psi|^{(n)}$ can be parameterized under the bases $\{\Omega_{i}\}^{d^2-1}_{i=0}$ as
\begin{equation*}\label{psi}
|\Psi\rangle\langle\Psi|^{(n)}=\frac{I}{d}+\sum^{d^2-1}_{i=1}\psi^{(n)}_i\Omega_i,
\end{equation*}
where $\psi^{(n)}_i=\textmd{Tr}(|\Psi\rangle\langle\Psi|^{(n)}\Omega_i)$.

When one performs measurements with measurement set $\{|\Psi\rangle\langle\Psi|^{(n)}\}^{M}_{n=1}$
on a collection of identically prepared copies of a quantum system (with state $\rho$), the probability to obtain the result of $|\Psi\rangle\langle\Psi|^{(n)}$ is
\begin{equation}\label{averageequation}
p_n=\textmd{Tr}(|\Psi\rangle\langle\Psi|^{(n)}\rho)
=\frac{1}{d}+\sum^{d^2-1}_{i=1}\Theta_i\psi_i^{(n)}\triangleq
\frac{1}{d}+\Theta^{\top}\Psi^{(n)}.
\end{equation}
Assume that the total number of experiments is $N$ and $N/M$ experiments are performed on $N/M$ identically prepared copies of a quantum system for each measurement basis $|\Psi\rangle\langle\Psi|^{(n)}$. Denote the corresponding outcomes as $x^{(n)}_1, \cdots, x^{(n)}_{N/M} $, which are independent and identically distributed. Let $\hat{p}_n=\frac{x^{(n)}_1+\cdots+ x^{(n)}_{N/M}}{N/M}$ and $e_n=\hat{p}_n-p_n$. According to the central limit theorem \cite{chow}, $e_n$ converges in distribution to a normal distribution with mean 0 and variance
$\frac{p_n-p_n^2}{N/M}$. Using (\ref{averageequation}), we have the linear regression equations for $n=1,\ 2,\ \cdots,\ M$,
\begin{equation}\label{average2}
\hat{p}_n=\frac{1}{d}+{\Psi^{(n)}}^{\top}\Theta+e_n,
\end{equation}
where $\top$ denotes the matrix transpose.

Note that the variance of $e_n$ is asymptotically $\frac{p_n-p_n^2}{N/M}$.  If $p_n=1$, we have already reconstructed the state as $|\Psi\rangle\langle\Psi|^{(n)}$; if $p_n=0$, we should choose the following measurement basis from the orthogonal complementary space of $|\Psi\rangle\langle\Psi|^{(n)}$.
$\hat{p}_n$, $d$ and $\Psi^{(n)}$ are all available for $n=1,\ \cdots,\ M$, while $e_n$ may be considered as the observation noise. Hence, the problem of quantum state tomography is converted into the estimation of the unknown vector $\Theta$. Denote $Y=\left(
           \begin{array}{ccc}
             \hat{p}_1-\frac{1}{d}, & \cdots, & \hat{p}_M-\frac{1}{d} \\
           \end{array}
         \right)^{\top}
$, $X=\left(
        \begin{array}{ccc}
          \Psi^{(1)}, & \cdots, & \Psi^{(M)} \\
        \end{array}
      \right)^{\top}
$, $e=\left(
           \begin{array}{ccc}
             e_1, & \cdots, & e_M \\
           \end{array}
         \right)^{\top}
$. We can transform the linear regression equations (\ref{average2}) into a compact form
\begin{equation}\label{average3}
Y=X\Theta+e.
\end{equation}

We define the MSE as E$\textmd{Tr}(\hat{\rho}-\rho)^2$, where $\hat{\rho}$ is an estimate of the quantum state $\rho$ based on the measurement outcomes and E$(\cdot)$ denotes the expectation on all possible measurement outcomes.
For a fixed tomography method, E$\textmd{Tr}(\hat{\rho}-\rho)^2$
depends on the state $\rho$ to be reconstructed and the chosen measurement bases.
From a practical viewpoint, the optimality of a chosen set of measurement bases
may rely upon a prior
information but should not depend on any specific unknown quantum state to be reconstructed.
In this Letter, no a prior assumption is made on the state $\rho$ to be reconstructed.
Given a fixed tomography method, we use
the maximum MSE  for all possible states (i.e., $\sup_{\rho}$E$\textmd{Tr}(\hat{\rho}-\rho)^2$) as the index
to evaluate the performance of a chosen set of measurement bases.
Hence, it is necessary to consider the worst case by enlarging the variance of the
observation noise $e_n$ in each linear regression equation. As a consequence,  $\{e_n\}^M_{n=1}$ may be treated as a set of independent identically distributed variables with asymptotic normal distribution $\text{N}(0, \frac{M}{4N})$. Another advantage of this treatment is that the effect of some other noises can be absorbed in the enlarged variance.

\emph{Asymptotic properties of the LS estimate}. To give an estimate with high accuracy and low computational complexity, we employ the LS method, where the basic idea is to find an estimate $\hat{\Theta}_{LS}$ such that
$$\hat{\Theta}_{LS}=\underset{\hat{\Theta}}{\text{argmin}}(Y-X\hat{\Theta})^{\top}(Y-X\hat{\Theta}),$$ where $\hat{\Theta}$ is an estimate of $\Theta$. Since the objective function is quadratic, one has the LS solution as follows:
\begin{equation}\label{ls1}
\hat{\Theta}_{LS}=(X^{\top}X)^{-1}X^{\top}Y=(X^{\top}X)^{-1}\sum^M_{n=1}\Psi^{(n)}(\hat{p}_n-\frac{1}{d}),
\end{equation}
where $X^{\top}X=\sum_{n=1}^M \Psi^{(n)}{\Psi^{(n)}}^{\top}.$

If the measurement bases $\{|\Psi\rangle\langle\Psi|^{(n)}\}^{M}_{n=1}$ are informationally complete or overcomplete, $X^{\top}X$ is invertible. Using (\ref{average3}), (\ref{ls1}) and the statistical property of the observation noise $\{e_n\}^M_{n=1}$ (asymptotically Gaussian), the estimate $\hat{\Theta}_{LS}$ has the following properties for a fixed set of chosen measurement bases:

1. $\hat{\Theta}_{LS}$ is asymptotically unbiased;

2. The MSE $\text{E}(\hat{\Theta}_{LS}-\Theta)^{\top}(\hat{\Theta}_{LS}-\Theta)$ of $\hat{\Theta}_{LS}$  is asymptotically  $\frac{M}{4N}\textmd{Tr}(X^{\top}X)^{-1}=\frac{M}{4N}\textmd{Tr}(\sum_{n=1}^M \Psi^{(n)}{\Psi^{(n)}}^{\top})^{-1}.$

3. $\hat{\Theta}_{LS}$ is asymptotically a maximum-likelihood estimate, and the estimation error can asymptotically achieve the Cram\'{e}r-Rao bound \cite{rao};

\emph{Positivity and computational complexity.} Based on the solution $\hat{\Theta}_{LS}$ obtained from (\ref{ls1}), we can obtain a Hermitian matrix $\hat{\mu}$ with $\textmd{Tr}\hat{\mu}=1$ using (\ref{rho}). However, $\hat{\mu}$ may have negative eigenvalues and be nonphysical due to the randomness of measurement results. In this sense,  $\hat{\mu}$ is called  pseudo linear regression estimation (PLRE) of state $\rho$. A good method of pulling $\hat{\mu}$ back to a physical state can reduce the MSE. In this Letter, the  physical estimate $\hat{\rho}$ is chosen to be the closest  density matrix to $\hat{\mu}$ under the matrix 2-norm. In standard state reconstruction algorithms, this task is computationally intensive \cite{smolin}. However, we can employ the fast algorithm in \cite{smolin} with computational complexity $O(d^3)$ to solve this problem since we have obtained a Hermitian estimate $\hat{\mu}$ with $\textmd{Tr}\hat{\mu}=1$.

Since an informationally complete measurement set $\{|\Psi\rangle\langle\Psi|^{(n)}\}^{M}_{n=1}$ requires $M$ being $O(d^2)$, the computational complexity of (\ref{rho}) and  $X^{\top}Y$  in (\ref{ls1}) is $O(d^4)$.  Although the computational complexity of calculating $(X^{\top}X)^{-1}$ is generally $O(d^6)$, $(X^{\top}X)^{-1}$ can be computed off-line before the experiment once the measurement  set is determined. Hence, the total computational complexity of LRE after the data have been collected is $O(d^4)$. It is worth pointing out that for $n$-qubit systems, $X^{\top}X=\sum_{n=1}^M \Psi^{(n)}{\Psi^{(n)}}^{\top}$ is diagonal for many preferred measurement sets such as tetrahedron and cube measurement sets. Fig.~1 compares the run time of our algorithm with that of a traditional MLE algorithm. Since the maximum MSE could reach 2 for the worst estimate, it is clear that  our algorithm LRE is much more efficient than MLE with a small amount of accuracy sacrificed.

\begin{figure}
\center{\includegraphics[scale=0.47]{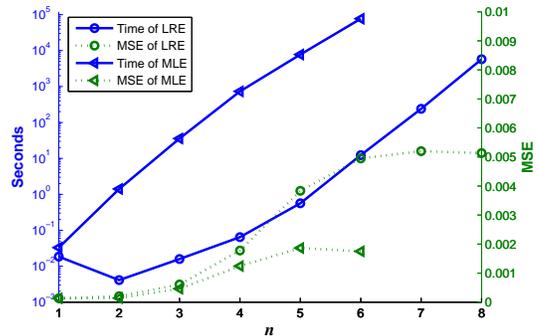}}
\caption{\label{computation time}The run time and MSE of LRE and MLE for  random  $n$-qubit  pure  states  mixed  with  the
identity \cite{smolin}. The realization of MLE used the iterative method in \cite{paris}. The measurement bases are from the $n$-qubit cube measurement set and the resource is $N=3^9\times4^n$.
The simulated measurement results for every base ${|\Psi\rangle\langle\Psi|}^{(i)}$ are generated from a binomial distribution with probability $p_i=\textmd{Tr}(|\Psi\rangle\langle\Psi|^{(i)}\rho)$  and trials $N/M$.
LRE is much more efficient than MLE with a small amount of accuracy sacrificed since the maximum MSE could reach 2 for the worst estimate. All timings were performed in MATLAB on the computer with
4 cores of 3GHz Intel i5-2320 CPUs. }
\end{figure}

\emph{Optimality of measurement bases}.  One of the advantages of LRE is that the MSE upper bound can be given analytically  as $\frac{M}{4N}\textmd{Tr}(\sum_{n=1}^M \Psi^{(n)}{\Psi^{(n)}}^{\top})^{-1}$, which is dependant explicitly upon the measurement bases. Note that if the PLRE $\hat{\mu}$ is a physical state, then the MSE upper bound is asymptotically tight for the evaluation of the performance of a fixed set of measurement bases. Hence, to choose an optimal set  $\{|\Psi\rangle\langle\Psi|^{(n)}\}^M_{n=1}$, one can solve the following optimization problem:
\begin{center}
Minimize  $\textmd{Tr}(\sum_{n=1}^M \Psi^{(n)}{\Psi^{(n)}}^{\top})^{-1}$\\
\ \ \ \ \ \ \   \   \   \   \    \ \ \ \ \ s.t.  ${\Psi^{(n)}}^{\top}\Psi^{(n)}=\frac{d-1}{d},$ for $n=1,\ \cdots,\ M.$
\end{center}
The optimization problem can be solved in an off-line way by employing appropriate algorithms though it may be computationally intensive \cite{optimal}.

With the help of the analytical MSE upper bound, we can ascertain which one is optimal among the available measurement sets. This is shown when we prove the optimality of several typical sets of measurement bases for 2-qubit systems below.

For 2-qubit systems, it is convenient to chose $\Omega_{i}=\frac{1}{\sqrt{2}}\sigma_l\otimes\frac{1}{\sqrt{2}}\sigma_m$, where $i=4l+m$;\ $l,\ m=0,\ 1,\ 2,\ 3$; $\sigma_0=I_{2\times2}$,
$\sigma_1=\left(
                                                                \begin{array}{cc}
                                                                  0 & 1 \\
                                                                1 & 0 \\
                                                               \end{array}
                                                              \right)
$, $\sigma_2=\left(
                                                               \begin{array}{cc}
                                                                0 & -i \\
                                                                 i & 0 \\
                                                               \end{array}
                                                             \right)
$, $\sigma_3=\left(
                                                                \begin{array}{cc}
                                                                  1 & 0 \\
                                                                  0 & -1 \\
                                                                \end{array}
                                                              \right)
$.
The MSE upper bound of  2-qubit states is
\begin{equation*}\label{2qubitMSE}
\frac{M}{4N}\textmd{Tr}(X^{\top}X)^{-1}=\frac{M}{4N}\textmd{Tr}(\sum^M_{n=1}\Psi^{(n)}{\Psi^{(n)}}^{\top})^{-1}.
\end{equation*}
Now we minimize this MSE upper bound or equivalently minimize $\textmd{Tr}(X^{\top}X)^{-1}$. Denote the eigenvalues of $X^{\top}X$ as $\lambda_1\geq\lambda_2\geq\cdots \geq\lambda_{15}$. Since we have $\psi^{(n)}_{0}=\frac{1}{2}$, $\sum^{15}_{i=0}{\psi^{(n)}_{i}}^2=1$, the subproblem is converted into
minimizing $\sum_{i=1}^{15} \frac{1}{\lambda_i}$, subject to $\sum_{i=1}^{15} \lambda_i=\frac{3}{4}M$. It can be proven that $\sum_{i=1}^{15} \frac{1}{\lambda_i}$ reaches its minimum $\frac{300}{M}$ when $\lambda_{1}=\cdots=\lambda_{15}=\frac{M}{20}$. Hence, the minimum of the MSE upper bound $\frac{M}{4N}\textmd{Tr}(X^{\top}X)^{-1}$ is $\frac{75}{N}$. This minimum MSE upper bound can be reached by using the mutually unbiased measurement bases.

If only local measurements can be performed, i.e., $|\Psi\rangle\langle\Psi|^{(n)}=|\Psi\rangle\langle\Psi|^{(n,1)}\otimes
|\Psi\rangle\langle\Psi|^{(n,2)},\ n=1,\ \cdots,\ M$, where
$|\Psi\rangle\langle\Psi|^{(n,1)}$ and $|\Psi\rangle\langle\Psi|^{(n,2)}$
can be parameterized as
$|\Psi\rangle\langle\Psi|^{(n,k)}=\sum^3_{l=0}\psi^{(n,k)}_l\frac{\sigma_l}{\sqrt{2}}$,\ $k=1,\ 2$. And we have $\psi^{(n)}_{i}=\psi^{(n,1)}_{l}\times\psi^{(n,2)}_{m}$, where $i=4l+m$. Due to additional constraints $\psi^{(n,k)}_0=\frac{1}{\sqrt{2}},\ \sum^{3}_{l=0} {\psi^{(n,k)}_l}^2=1,$ for $k=1,\ 2$, $n=1,\ \cdots,\ M$, the subproblem of minimizing the MSE upper bound can be converted into minimizing
$\sum_{i=1}^{15} \frac{1}{\lambda_i}$, subject to (i) $\sum_{i=1}^{3} \lambda_i\geq \frac{1}{4}M$; (ii) $\sum_{i=1}^{6} \lambda_i\geq \frac{1}{2}M$; (iii) $\sum_{i=1}^{15} \lambda_i=\frac{3}{4}M$. It can be proven that $\sum_{i=1}^{15} \frac{1}{\lambda_i}$ reaches its minimum $\frac{396}{M}$ when $\lambda_{1}=\cdots=\lambda_{6}=\frac{M}{12}$, $\lambda_{7}=\cdots=\lambda_{15}=\frac{M}{36}$.
Hence, the minimum of the MSE upper bound $\frac{M}{4N}\textmd{Tr}(X^{\top}X)^{-1}$ is $\frac{99}{N}$. This minimum MSE upper bound can be reached by using the 2-qubit cube or tetrahedron measurement set.

\begin{figure}
\center{\includegraphics[scale=0.50]{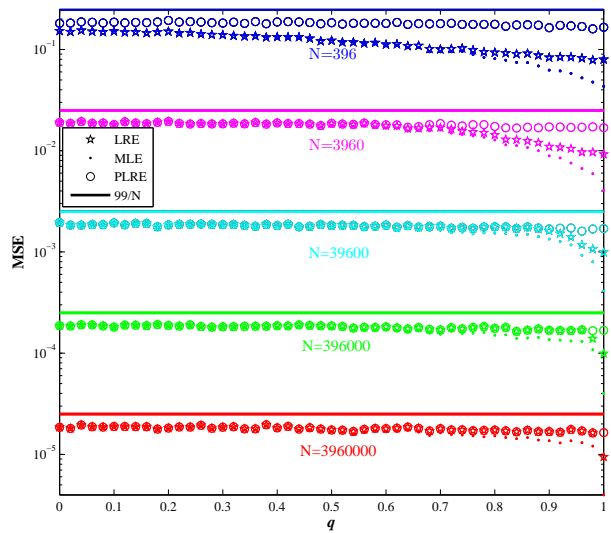}}
\caption{\label{LREvsMLE} Mean squared error (MSE) for Werner states \cite{wernerstate} with $q$ (varying from 0 to 1) and different numbers of copies $N$. The cube measurement set is used, where the MSE upper bound is $\frac{99}{N}$.  It can be seen that the MSE of PLRE is almost unchanged for $q\in[0,1]$, and is larger than the  MSE of LRE.}
\end{figure}

Fig. 2 shows the dependant relationships of the MSEs for Werner states \cite{wernerstate}
on $q$ (varying from 0 to 1) and different number of copies $N$ using the
cube measurement bases \cite{adamson}. The fact that the MSE of PLRE is larger than
that of LRE demonstrates that the process of pulling $\hat{\mu}$ back to a physical
state further reduces the estimation error.

\emph{Discussions and conclusions}.
In the LRE method, data collection is achieved by performing measurements on quantum systems with given measurement bases. This process can also be accomplished by considering the evolution of quantum systems with fewer measurement bases. For example, suppose only one observable $\sigma$ is given, and the system evolves according to a unitary group $\{U_t\}$. At a given time $t$,
\begin{equation*}\label{sigma}
\langle\sigma_t\rangle=\textmd{Tr}(U^\dag(t)\sigma U(t)\rho)=\textmd{Tr}(\sigma_t\rho).
\end{equation*}
Suppose one measures the observable $\sigma$ at time $t$ ($t=1,\ \cdots,\ M$) on $m$ identically prepared copies of a quantum system. Denote the obtained outcomes as $\sigma^t_1,\ \cdots,\ \sigma^t_{m}$, and their algebraic average as $\bar{\sigma}_t=\frac{\sigma^t_1+\cdots+\sigma^t_{m}}{m}$. Note that $\sigma^t_1,\ \cdots,\ \sigma^t_{m}$ are independent and identically distributed. According to the central limit theorem \cite{chow}, $e_t=\bar{\sigma}_t-\langle\sigma_t\rangle$ converges in distribution to a  normal distribution with mean 0 and variance $\frac{\langle\sigma^2_t\rangle-\langle\sigma_t\rangle^2}{m}$.  We have the following linear regression equations
\begin{equation*}
\bar{\sigma}_t=\textmd{Tr}(\sigma_t\rho)+e_t,\  \ \ \     t=1,\ \cdots,\ M,
\end{equation*}
which are similar to (\ref{average2}). Hence, we can use the proposed LRE method to accomplish quantum state tomography.

The LRE method can also be extended to reconstruct quantum states with a prior information \cite{cramer,klimov,Gross2010,toth2010} or states of open quantum systems. Actually, LRE can be applied whenever there are measurable quantities that are linearly related to all density matrix elements of the quantum system under consideration.

In conclusion, an efficient method of linear regression estimation has been presented for quantum state tomography. The computational complexity of LRE is $O(d^4)$, which is much lower than that of MLE and BME. We have analytically  provided an  MSE upper bound for all possible states to be estimated, which explicitly depends upon the used measurement bases. This analytical upper bound can assist to identify optimal measurement sets. The LRE method has potential for wide applications in real experiments.

The authors would like to thank Lei Guo, Huangjun Zhu and Chuanfeng Li for helpful discussion. The work in USTC is supported by National Fundamental Research Program (Grants No. 2011CBA00200 and No. 2011CB9211200), National Natural Science Foundation of China (Grants No. 61108009 and No. 61222504), Anhui Provincial Natural Science Foundation(No. 1208085QA08). B. Q. acknowledges the support of National Natural Science Foundation of China (Grants No. 61004049, No. 61227902 and No. 61134008). D. D. is supported by the Australian Research Council (DP130101658).


\begin{thebibliography}{99}
\bibitem{optimal}
We will dicuss this problem in other work.
\end{thebibliography}
\end{document}